\documentclass[review]{elsarticle}

\usepackage{lineno,hyperref}
\usepackage{graphicx,color}
\usepackage{sidecap}
\usepackage{amsmath}
\modulolinenumbers[5]

\journal{Advances in Magnetism at the Joint European Magnetic Symposia 2018 (JEMS2018)}

\newcommand{\Cu}{Cu$_4$Cu}

\newcommand{\Co}{Co$_4$Co}
\begin{document}
\begin{frontmatter}
\title{High-frequency EPR study on {\Cu}- and {\Co}-metallacrown complexes}

\author[KIP]{C.~Koo\corref{mycorrespondingauthor}}
\cortext[mycorrespondingauthor]{Corresponding author}
\ead{changhyun.koo@kip.uni-heidelberg.de}
\author[KIP]{J.~Park}
\author[KIP]{J.~Butscher}
\author[Mainz]{E.~Rentschler}
\author[KIP,CAM]{R.~Klingeler}

%\author[Mainz]{P.~Happ}

\address[KIP]{Kirchhoff Institute of Physics, Heidelberg University, INF 227, D-69120 Heidelberg, Germany}
\address[Mainz]{Institut f\"{u}r Anorganische und Analytische Chemie, Universit\"{a}t Mainz, D-55128 Mainz, Germany}
\address[CAM]{Centre for Advanced Materials, Heidelberg University, INF 225, D-69120 Heidelberg, Germany}

%%%%%%%%%%%%%%%%%%%%%%%

\sloppy

\begin{abstract}
High-frequency/high-field electron paramagnetic resonance studies on two homo\-nuclear 12-MC-4 metallacrown complexes \Cu\ and \Co\ are presented. For \Cu , our data imply axial-type $g$-anisotropy with $g_{x}$ = 2.03 $\pm$ 0.01,  $g_{y}$ = 2.04 $\pm$ 0.01, and $g_{z}$ = 2.23 $\pm$ 0.01, yielding $g=2.10 \pm 0.02$. No significant zero field splitting (ZFS) of the ground state mode is observed. In \Co , we find a m$_S$ = $\pm$3/2 ground state with $g=2.66$. The data suggest large anisotropy $D$ of negative sign.
\end{abstract}

\begin{keyword}
MOF, magnetism, magnetic anisotropy, high-frequency electron paramagnetic resonance
\end{keyword}
\end{frontmatter}
%\linenumbers

\section{Introduction and Experiment}

Metallacrowns are coordination compounds including metal ions where a repeating sequence, i.e., [--M--O--N--], forms macroscopic rings.~\cite{Happ2014} The characteristic rings of regular metallacrowns involve oxygens pointing towards the core of the cycle, thereby offering coordination positions of central cations. The class of 12-MC-4 studied at hand comprises 12 atoms in the planar square cycle, i.e., four of which being metal ions, with an additional metal ion in slightly off-plane center site. Here, we present high-frequency/high-field electron paramagnetic resonance (HF-EPR) studies on two metallacrown 12-MC-4 complexes: (HNEt$_3$)$_2$Cu(II)[12-MC$_{\rm Cu(II)N(Shi)}$-4] (i.e., \Cu ) and the mixed-valent (HPip)(Piv)[Li[Co(II)($\mu$2-Piv)$_2$(Piv)[12-MC$_{\rm Co(III)N(Shi)}$-4](Pip)$_5$]]$_2$) (i.e., \Co ). In \Co , the low-spin Co(III) ions which are diamagnetic form the scaffold while the Co(II)-ion is located in the center position. Details of the synthesis method, structure information, and characterisation are reported in Ref.~\cite{Happ2014,Happ2014Fe4Cu}.

%\Fe4Cu\ complex ( Cu(II)Cl$_2$(DMF)$_2$[12-MC$_{\rm Fe(III)N(Shi)}$-4](DMF)$_4$.

%\section{Experimental}

HF-EPR measurements were carried out using a phase-sensitive millimeter-wave vector network analyzer (MVNA) from AB Millim\`{e}tre covering the frequency range from 30 to 1000 GHz~\cite{Comba2015}. For each frequency range (Q,L,W band etc.), different sets of Schottky diode systems were used. Experiments were performed in a 16 T superconducting magnet with temperature control sensors in both probe and sample space. Loose powder was densely packed in the sample space of the cylindrical waveguide probe without any glue or grease. Analysis of the obtained EPR experimental data were performed using the program EasySpin~\cite{EasySpin}.

%\section{Results and Discussion}

\subsection{Results on \Cu}

\begin{figure}
\includegraphics[width=0.95\columnwidth,clip] {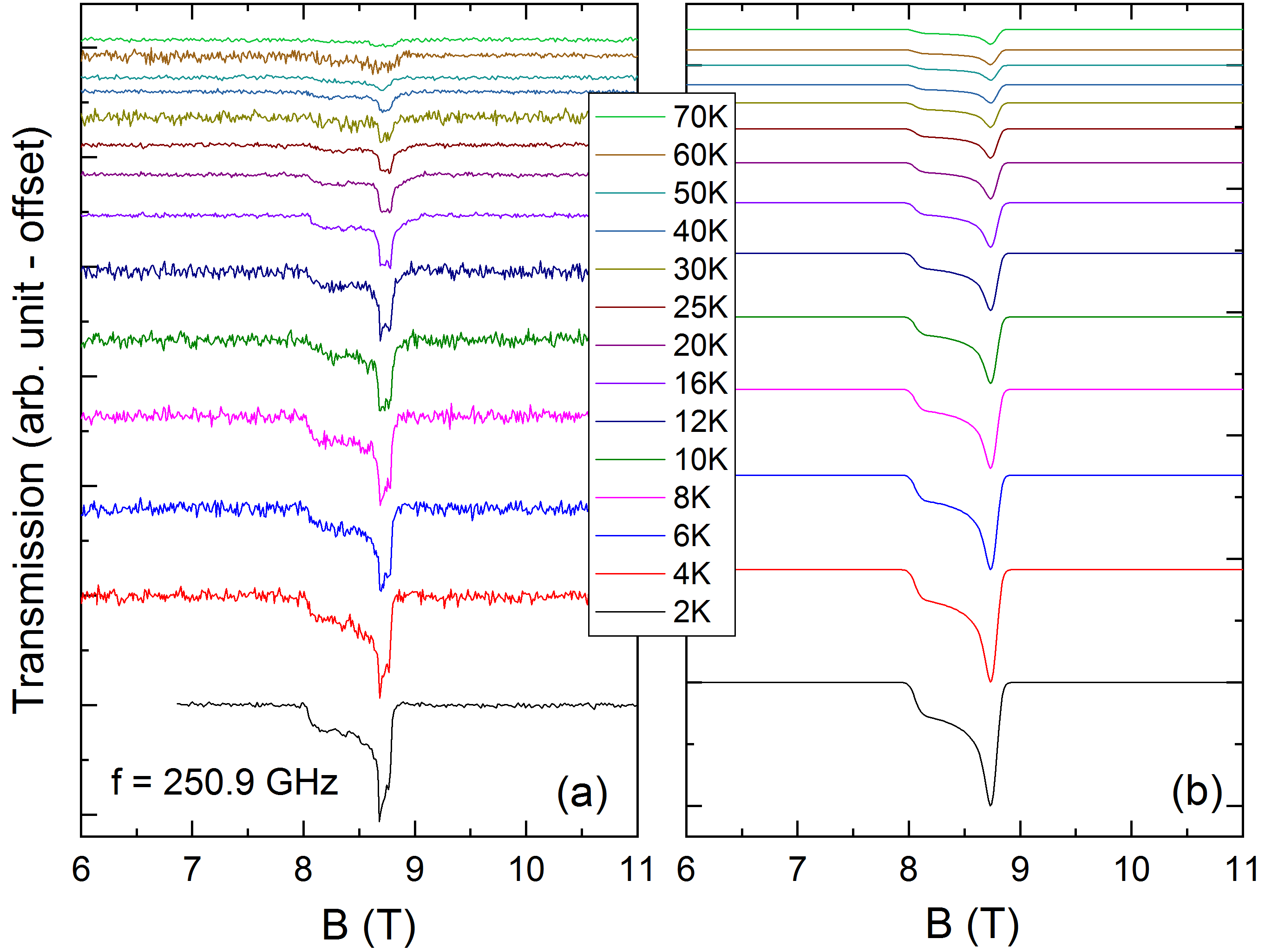}
\caption{HF-EPR spectra of \Cu\ at various temperatures and $f=250.9$~GHz. (a) Experimental spectra, and (b) simulated spectra.}\label{fig:Cu5-Temp}
\end{figure}

HF-EPR spectra of \Cu\ obtained at $f=250.9$~GHz are shown in Fig.~\ref{fig:Cu5-Temp}. We observe one broad resonance feature which implies the presence of a S = 1/2 ground state doublet in agreement of the susceptibility data analysis \cite{Happ2014Fe4Cu}. The resonances resemble typical powder spectra~\cite{Koo2016}. At $T=2$~K, a sharp peak appears in the high-field region of the resonance accompanied by a broad shoulder at lower field. This shape of the resonance clearly indicates an easy axis-type of $g$-anisotropy in the complex. The temperature dependence of the spectra at $f=250.9$~GHz shows a Curie-like decrease upon heating. Simulation of the resonances in terms of a uniaxial magnetic system with two magnetic principal axes corresponding to three components of the $g$-tensor yields a good description of the data as displayed by the experimental (a) and simulated (b) data in Fig.~\ref{fig:Cu5-Temp}. Note, however, that the data exhibit a broad double feature of the resonance which is not reproduced by the simulation. We attribute this either to not fully aligned powder or to a few impurity spins in the complex.

\begin{figure}
\includegraphics[width=0.95\columnwidth,clip] {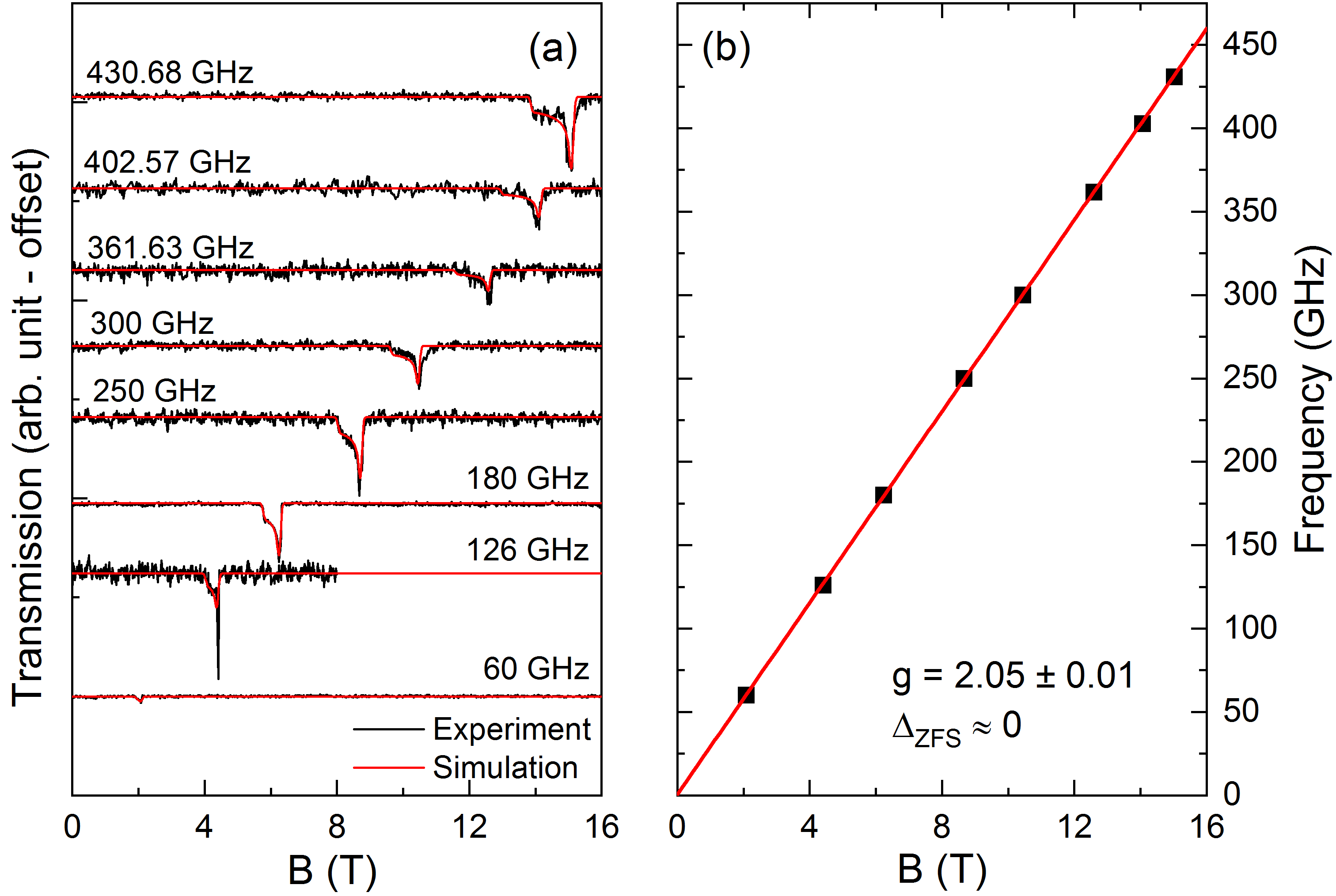}
\caption{(a) HF-EPR spectra of \Cu\ at various frequencies and at $T=4$~K, and (b) frequencies vs. resonance fields. Black lines and symbols represent experimental data, and red lines represent simulations or fits to the data}\label{fig:Cu5-Freq}
\end{figure}

Quantitatively, the simulation yields g$_{x}$ = 2.03 $\pm$ 0.01, g$_{y}$ = 2.04 $\pm$ 0.01, and g$_{z}$ = 2.23 $\pm$ 0.01. As the sharp feature in the spectra corresponds to $g_z$, it can be determed from the spectra at various frequencies and at $T=4$~K presented in Fig.~\ref{fig:Cu5-Freq}a. Again, all spectra are well simulated with the identical powder spectra simulation parameters. The resonance fields at the minima of the sharp resonance features linearly depend on the microwave frequency (see Fig.~\ref{fig:Cu5-Freq}b). While the associated slope confirms the value of $g_z$, it also shows that the zero-field splitting associated with the resonance is negligible.

\subsection{Results on \Co }

\begin{figure}[h]
\center{\includegraphics [width=0.95\columnwidth,clip] {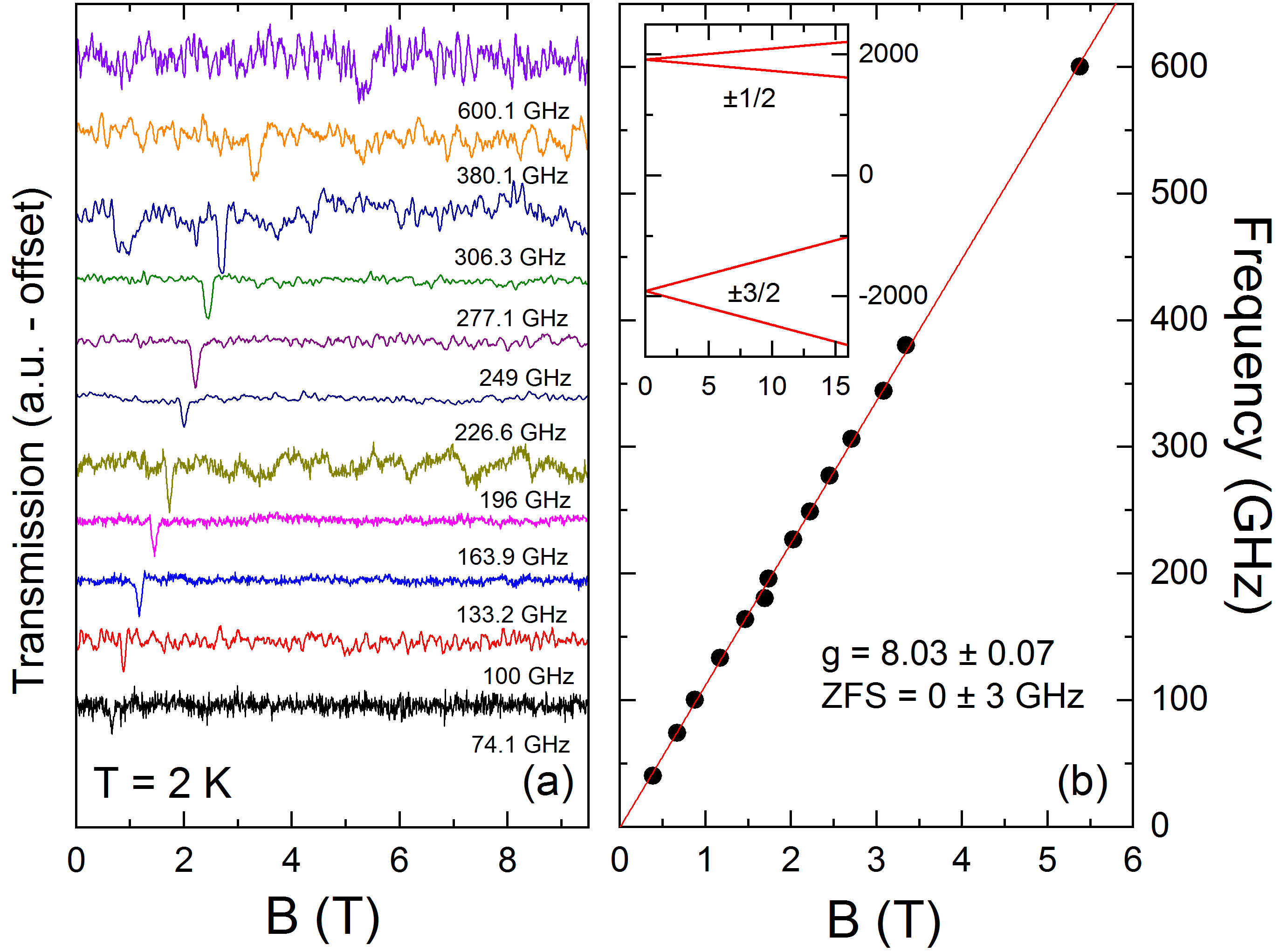}}
\caption[] {\label{cof}(a) HF-EPR spectra and (b) frequency vs. resonance field diagram of \Co\ at $T=2$~K. The line shows a linear fit to the data and the inset is a simulation of the energy level diagram (see the text).}
\end{figure}

The HF-EPR spectra, at $T=2$~K, on \Co\ display a single peak feature (Fig.~\ref{cof}a). The frequency dependence of the resonance fields in Fig.~\ref{cof}b shows linear behaviour. The slope of the resonance branch corresponds to a $g$-factor of $g_{\rm fit}$ = $g\cdot\Delta$m$_S$ = 8.03(7) implying a forbidden resonance, $\Delta$m$_S$ $\neq \pm$1 with g = 2.00. Considering the metallacrown arrangement of the Co(II) ion hosted in a strongly distorted octahedron in the center of four diamagnetic Co(III)-ions suggests to attribute the resonance to the paramagnetic center ion.~\cite{Happ2014} From the observed slope of the resonance branch, we conclude that we observe a transition from the m$_S$ = -3/2 to the m$_S$ = 3/2 spin state which is allowed with the spin state mixing due to the off-axial anisotropy. Under this assumption, the effective $g$-factor amounts to $g = 2.66(2)$. This is in good accordance to reported Co(II)-ions in distorted octahedral environment.~\cite{Shin, Polunin, Mueller} It also agrees to the $\chi T$-curve that obeys the characteristic behavior of an isolated six-coordinate Co(II) ion. From the high-temperature value of $\chi T$ the effective $g$-factor of 2.60 had been concluded previously~\cite{Happ2014} which reasonably matches to the more precise $g$-value derived from the EPR study at hand.

\begin{figure}[h]
\center{\includegraphics [width=0.8\columnwidth,clip] {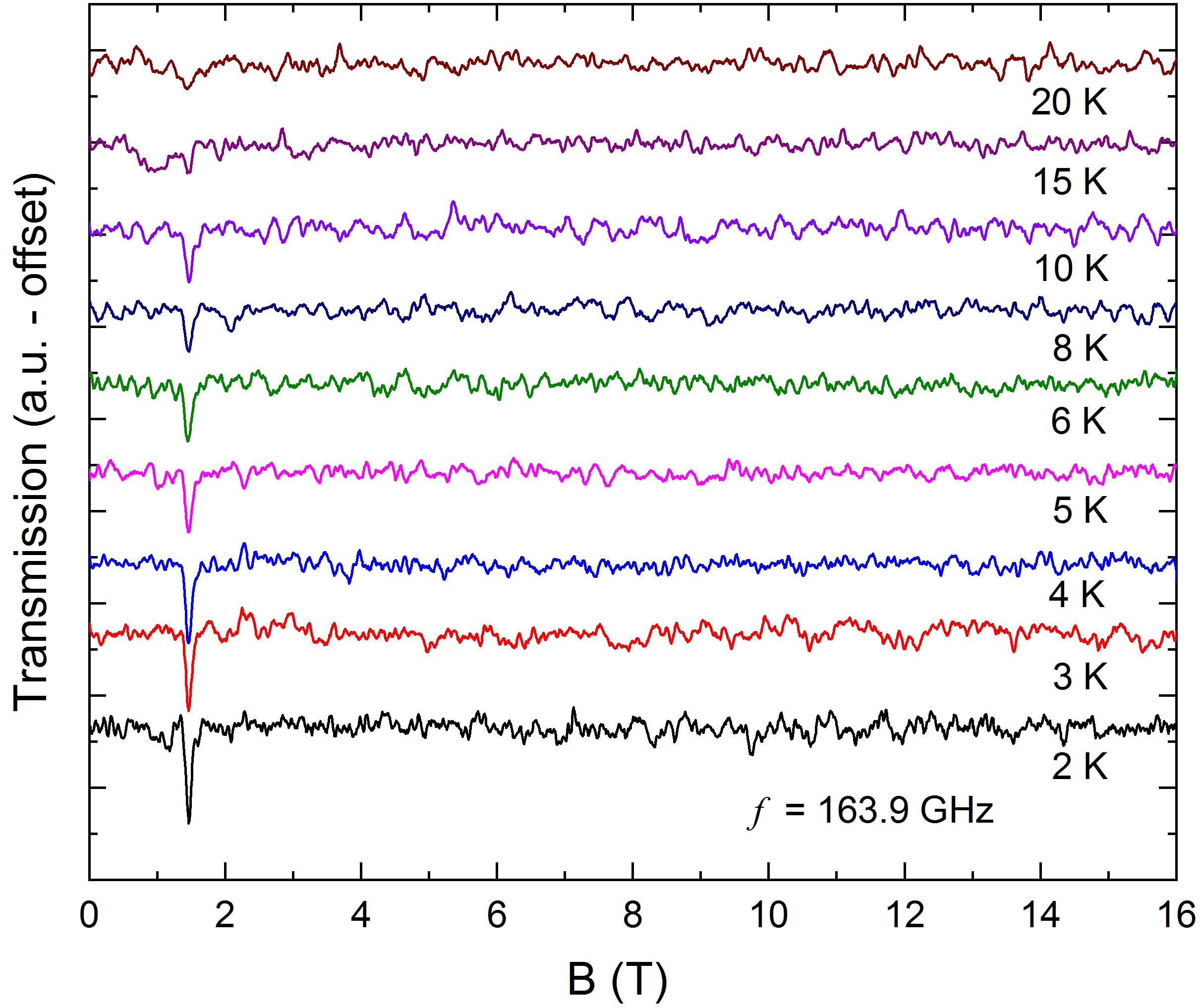}}
\caption[] {\label{cot}HF-EPR spectra of \Co\ at constant frequency $\nu =163.9$ GHz and different temperatures. The visible peak vanishes at about 20~K.}
\end{figure}

As expected for the $\Delta$m$_S$ = 3 transition in the m$_S$ = $\pm$ 3/2 Kramer's doublet spin state, negligible zero-field splitting is observed in the experimental data. The temperature dependence of the HF-EPR spectra at 163.9~GHz presented in Fig.~\ref{cot} shows a Curie-like decrease of intensity of the resonance feature upon heating and its disappearance at around 20 K. Note, that no additional, i.e., thermally activated resonances are observed up to 50~K. This is explained by large magnetic anisotropy $D$. In Ref.~\cite{Happ2014} it is argued that anisotropy of \Co\ is larger than $D = -64$~cm$^{-1}$. Similarly large values are found in the literature for isolated high-spin Co(II) complexes in distorted octahedral coordination where, e.g., $g=2.580$, $|D| = 87.9$~cm$^{-1}$ \cite{Shin}, $|D| = 60(3)$~cm$^{-1}$, $g_z = 2.77(5)$, $g_{xy}=3.04(5)$ \cite{Polunin}, and $|D_1| = 63(1)$~cm$^{-1}$, $|D_2| = 58(1)$~cm$^{-1}$,$g_1=2.53(1)$, $g_2=2.56(1)$ \cite{Mueller}. In order to assess the energy level diagram we used a phenomomenological Hamiltonian consisting of a Zeeman and a zero field splitting term, which is given by

\begin{equation}
\hat{H}=g\mu_BS_zB+D\left( S_z^2+\frac{S(S+1)}{3}\right).
\end{equation}

The resulting simulated diagram using $g = 2.66$ and $D = -64$~cm$^{-1}$ is shown as inset in Fig~\ref{cof}b. It illustrates that the excited $\pm 1/2$ states are separated by nearly 200~K from the $\pm 3/2$ ground states.

\section{Discussion and summary}

Powder HF-EPR spectra presented here confirm the $S=1/2$ ground state in \Cu . From previous susceptibility data, the exchange couplings $J_1=−155.2$~cm$^{−1}$ between the center and corner spins, and $J_2=−92.3$~cm$^{−1}$ between the nearest neighbor corner spins as well as $g = 2.16$ had been deduced~\cite{FN1,Happ2014}. The HF-EPR data imply axial-type $g$-anisotropy with $g_{x}$ = 2.03 $\pm$ 0.01,  $g_{y}$ = 2.04 $\pm$ 0.01, and $g_{z}$ = 2.23 $\pm$ 0.01, yielding $g=(g_x+g_y+g_z)/3=2.10 \pm 0.02$. This value only slightly disagrees to the one observed in the static susceptibility. As expected for a Cu(II) system, neither significant zero field splitting (ZFS) of the ground state mode nor excited states are observed. This is in accordance to the minimal model using $J_1$ and $J_2$ mentioned above which implies the $S=1/2$ ground state being well separated from excited and $S=3/2$ states by 125.8 and 184.5~cm$^{−1}$ ($S=1/2$) and 213.8~cm$^{−1}$ ($S=3/2$).~\cite{Happ2014Fe4Cu} In \Co , we find a $\pm 3/2$ ground state with $g=2.66$. The fact that the $\pm 3/2$ states are favoured with respect to the $\pm 1/2$ states implies negative sign of anisotropy $D$. Upon heating to 50~K we do not observe any signatures of excited state resonances which is consistent to a large values of $|D|$ of supposingly about 200~K. The data at hand demonstrate, that HF-EPR yields precise information on the actual spin ground state, the $g$-factor, and magnetic anisotropy in transition metal coordination complexes.

%In \fecu , the spins $S_{\rm Cu}=1/2$ and $S_{\rm Fe}=5/2$ form an intermediate spin $S=11/2$ ground state.~\cite{Happ2014Fe4Cu}. The ground state resonances can be straightforwardly attributed to the main (black) resonance in Figs.~\ref{Fe4CuT} and \ref{Fe4CuB}. The next excited states are $S=13/2$ and $S=9/2$ which are only 4.6 and 5.9~K higher in energy and may well be assigned to the (blue and red) excited state features found in HF-EPR. Even small temperatures are sufficient to populate these low-lysing excited state, yielding thermally activated EPR resonance lines.~\cite{Golze}

%It has been pointed out very recently, however, that determining magnetic interaction and the ground state in the case of nearly degenerated states at hand by means of analysis of static susceptibility data is not unique.~\cite{Pavl2018} This particularly holds for the fact that the respective states exhibit significant anisotropy yields different zero field splitting of the ground state and the excited ones.


\begin{thebibliography}{10}
\bibitem{FN1} Note, that the exchange integrals refer to the Hamiltonian $\sum$ -2$\cdot$J$_{ij}\hat{S}_i\hat{S}_j$.
\end{thebibliography}
\end{document}